\documentclass{PoS} 
\usepackage{epsfig}

\title{
\vspace*{-2.6cm}
\begin{minipage}{\textwidth}
{\normalfont\small LTH 879, DESY 10-118
\hspace{\fill} August 2010}\\
\end{minipage}\\[35pt]
Threshold-improved predictions for charm production in 
deep-inelastic scattering}

\ShortTitle{Threshold improved predictions for charm production in DIS}

\author{\speaker{N. A. Lo Presti}\\
        Department of Mathematical Sciences, University of Liverpool, UK\\
        E-mail: \email{lopresti@liv.ac.uk}}

\author{H. Kawamura$\,$\thanks
        {Present address: KEK Theory Center, Tsukuba, Japan}\\
        Department of Mathematical Sciences, University of Liverpool, UK\\
        E-mail: \email{kawamura@liv.ac.uk}}
        
\author{S. Moch\\
        Deutsches Elektronensynchrotron DESY, Zeuthen, Germany\\
        E-mail: \email{sven-olaf.moch@desy.de}}
        
\author{A. Vogt\\
        Department of Mathematical Sciences, University of Liverpool, UK\\
        E-mail: \email{Andreas.Vogt@liv.ac.uk}}

\abstract{
\vspace*{1cm}
We have extended previous results on the threshold expansion of the gluon 
coefficient function for the charm contribution to the deep-inelastic structure
function $F_2$ by deriving all threshold-enhanced contributions at the 
next-to-next-to-leading order. The size of these corrections is briefly 
illustrated, and a first step towards extending this improvement to more 
differential charm-production cross sections is presented.
}

\FullConference{XVIII International Workshop on Deep-Inelastic Scattering and Related Subjects\\
		 April 19-23, 2010, Convitto della Calza, Firenze, Italy}

\newcommand{\gsim}{\raisebox{-0.07cm}{$\:\:\stackrel{>}{{\scriptstyle
 \sim}}\:\: $} }
\newcommand{\lsim}{\raisebox{-0.07cm}{$\:\:\stackrel{<}{{\scriptstyle
 \sim}}\:\: $} }

\newcommand{\hspp}{{\hspace{8mm}}}
\newcommand{\beq}{\begin{equation}}
\newcommand{\eeq}{\end{equation}}
\newcommand{\bea}{\begin{eqnarray}}
\newcommand{\eea}{\end{eqnarray}}
\newcommand{\nn}{\nonumber}
\newcommand{\ra}{\rightarrow}

\newcommand{\MSb}{$\overline{\mbox{MS}}$}
\newcommand{\GeVs}{\mbox{GeV}^{\:\!2}}
\newcommand{\Qs}{Q^{\:\!2}}
\newcommand{\ms}{m^{\:\!2}}
\newcommand{\as}{\alpha_{\sf s}}

\def\nf{{n^{}_{\! f}}}
		 
\begin{document}

\section{Introduction}
 
\noindent
Deep-inelastic scattering (DIS), measured in fixed-target experiments and at
HERA, provides core constraints on the parton distributions for the LHC. 
For some crucial processes, such as gauge-boson and Higgs production, these
distributions are required at the next-to-next-to-leading order (NNLO) of
perturbative QCD. Consequently coefficient functions at this accuracy are 
needed also for the extraction of the parton densities from (mainly) the 
structure function $F_2(x,\Qs)$ in DIS. 

\vspace{1mm}
For the massless case, these quantities have been known for a long time 
\cite{c2light}. 
However, a considerable part of $F_2$ at small Bjorken-$x$ is due to the 
production of charm quarks which is dominated by the photon-gluon fusion 
process $\gamma^{\,\ast}g \ra c \bar{c}\,X$. The NLO coefficient functions for 
$F_2^{\:\!c}$ have been obtained in a semi-analytic manner \cite{LRSvN93}; 
the results are often used via the parametrizations of Ref.~\cite{RSvN94}
(for minor corrections see Ref.~\cite{HS95}). 
The corresponding NNLO corrections are not known.
Fully analytic NLO results have obtained in the asymptotic limit $m_c^{\,2}/\Qs 
\ra 0$ \cite{F2c-asympt}. Recently these calculations have been extended to
NNLO for the lowest even-integer Mellin moments \cite{BBK09}.

\vspace{1mm}
It has been known for a long time, see, e.g., Refs.~\cite{GRS+AV}, that, at not
too large values of $\Qs$, the convolution of the coefficient function for 
$F_2^{\:\!c}$ and the gluon density is dominated by rather low partonic 
of-mass energies (CM). Hence the NNLO predictions of the threshold resummation 
\cite{SoftGlue,SoftglueH} can provide useful information on the dominant 
contribution to $F_2^{\:\!c}$.
Previously the first two \cite{LM98} and three \cite{AM08} highest
threshold logarithms have been determined at this and all higher orders. 
 
\vspace{1mm}
In this contribution we employ recent developments concerning the structure of 
massive-particle amplitudes and the description of heavy-quark production in 
hadronic collisions \cite{NeubertXX,BenekeXX} to extend those results to four 
logarithms, i.e., we are now able to derive all threshold-enhanced terms at 
NNLO. We also include a brief update of the results of Ref.~\cite{LM98} for
the transverse momentum distributions, calculated at NLO in Refs.~\cite
{LRSvN93a}, using a modern set of parton distributions \cite{ABKM09}.

\section{Threshold resummation of the gluon coefficient function for 
$F_2^{\:\!c}$}

\noindent
The heavy-quark coefficient functions for $F_2$ are usually expressed in terms 
of the variables
\beq
\label{variables}
  \xi   \:=\: \frac{\Qs}{\ms}    \;\; , \;\; \mbox{ and } \;\;\; 
  \eta  \:=\: \frac{1}{\rho} - 1 \;\;\; \mbox{ or } \;\;\;
  \beta \:=\: \sqrt{1 - \rho}    \;\;\; \mbox{ with } \;\;\;
  \rho  \:=\: \frac{4\,\ms}{s}   \;\; ,  
\eeq
where $s$ is the CM energy, $m$ the mass of the heavy quark, and $\beta$ the
relative velocity of the heavy-quark pair.
In terms of the threshold limit, $\rho$ corresponds to the Bjorken variable $x$
in massless DIS. Hence the dominant gluon coefficient function receives a 
double-logarithmic higher-order enhancement at $\beta \ll 1$. 
The resummation of these logarithms is performed in terms of the Mellin 
variable $N$ conjugate to $\rho$. Up to terms suppressed by powers of $1/N$, 
the coefficient function reads
\beq
\label{exponentiation}
  c_{2,g}^{} \,(\as, N) \;\;=\;\; 
  c^{\,(0)}_{2,g}(N)\,\cdot\, g^{}_0(\as, N) \,\cdot \,
  \exp \left[ \,G (\as, \ln N) \, \right] \;\; .
\eeq
Here $c^{\,(0)}_{2,g}$ is the lowest-order coefficient function (see, e.g.,
Ref.~\cite{RSvN94}), and $g^{}_0(\as, N)$ a matching coefficient. Its 
dependence on $N$, absent in the massless case, is due to Coulomb terms which 
are enhanced by a factor $1/\beta$ (see below). 
The resummation exponent $G$ is of the standard form
\beq
\label{exp:exponent}
  G \;\;=\;\;  
  \int_0^1 dz\,\frac{z^{\,N-1} -1}{1-z} \:
  \bigg[
  \int_{\mu^2}^{\,4m^2(1-z)^2}
    \frac{dq^2}{q^2}\: A_g(\as (q^2\:\!) ) 
    \:+\: D_{\gamma^{\,\ast} g\rightarrow c\bar{c}}\,(\as (4m^2[1-z]^2\,) )
  \bigg] \;\; .
\eeq
The first term resums the collinear gluons emitted off the initial gluon, the
corresponding `cusp anomalous dimension' $A_g$ is known to order $\as^3$ 
\cite{MVV4}. The second term collects soft and final-state emissions. 
Following the methods of Refs.~\cite{NeubertXX,BenekeXX} we find
\beq
\label{Dcoeff}
   D_{\gamma^{\,\ast} g\rightarrow c\bar{c}} \;\; = \;\;
   1/2\: D_{gg \ra Higgs} \:+\: D_{Q\bar{Q}} \;\; ,
\eeq
where the latter heavy-quark coefficient is known to order $\as^2$ 
\cite{BenekeXX} (obviously only the colour-octet result is required in the 
present case), and the former even to order $\as^3$ \cite{D-Higgs}.

\vspace{1mm}
The above information is sufficient to predict the highest four powers of 
$\ln N$ at all orders in~$\as$ (cf., e.g., Ref.~\cite{MVV7}), provided that 
the matching function $g_0$ is known at NLO. $g_0$ is of the form
\beq
\label{g0coeff}
   g_0(\alpha_s, N) \;\;=\;\; 
   g_0^{\,h}(\alpha_s)\;\cdot\; g_0^{\,c}(\alpha_s, N) \:\: .
\eeq
The Coulomb contribution $g_0^{\,c}$ can be determined by Mellin transforming 
the partonic cross section in non-relativistic QCD, calculated for the 
colour-singlet case to NNLO in Ref.~\cite{Czarnecki:2001gi}. The required octet
results are obtained by the colour-factor replacement $\,C_F \ra C_F-C_A/2\,$.
The NLO contribution to the $N$-independent hard matching constant 
$g_0^{\,h}$ had not been determined before this research. 
We have extracted this coefficient -- which will be presented elsewhere 
\cite{lPKMVprep} -- analytically by integrating the intermediate results of 
Ref.~\cite{LRSvN93} (distributed as a {\sc Fortran} program), and checked our 
result numerically, for some relevant values of $\xi$, using the 
parametrization of Ref.~\cite{RSvN94}.
	
\section{Threshold approximation to the NNLO coefficient function}

\noindent
The above $N$-space results can be readily expanded in $\as$ and then Mellin 
inverted using, e.g., App.~A of Ref.~\cite{MochUwer08} and the fact that the 
leading-order coefficient function is linear in $\beta$ near threshold
(we normalize the coefficient functions as in Refs.~\cite{LRSvN93,RSvN94}),
\beq
\label{c0thresh}
   c_{2,g}^{(0)}(\xi, \beta) \;\; = \;\;  \pi T_f \: \beta\, (1\,+\,\xi/4)^{-1} 
   \:+\: \mathcal{O}(\beta^3) \;\; .
\eeq
At NLO one thus recovers the threshold expansion (with $T_f = 1/2$, $C_A = 3$
and $C_F = 4/3$ in QCD)
\bea
\label{c1thresh}
  c_{2,g}^{\,(1)}(\xi, \beta) &\:\: = \:\:&
  \frac{c_{2,g}^{(0)}}{(4\pi)^2} \:
  \Big\{ 4\,C_A \,\ln^2 (8\beta^2) - 20\,C_A  \ln(8\beta^2) + \,c_0(\xi)\,  
	+ (2\,C_F - C_A)\: \frac{\pi^2}{\beta} \nonumber\\[-1.5mm]   
&& \mbox{\hspp}
  +\ln\frac{\mu^2}{\ms} \, \big[ - 4\,C_A  \ln(4\beta^2)+ 
  \,\bar{c}_0(\xi)\,\big] + \mathcal{O}(\beta^2) \bigg\} \;\; .
\eea
The logarithmic and $1/\beta$ contributions have first been given in
Ref.~\cite{RSvN94}. The scale term $\bar{c}^{}_0(\xi)$ is fixed by 
renormalization-group constraints and reads
\beq
\label{c0bar}
  \bar{c}^{}_0(\xi) \;\; = \;\; 
  4\,C_A \,(\, 2+\ln(1 + \xi/4) ) \: - \: 4/3\: T_f
  \;\; ,
\eeq
where the final term arises from the transformation of $\as$ to the standard 
\MSb\ scheme \cite{BBK09plb} which was not performed in Ref.~\cite{RSvN94}.
The corresponding scale-independent contribution $c^{}_0(\xi)$ is not available
in the literature yet, the full result will be presented in 
Ref.~\cite{lPKMVprep}. Here we can, for brevity, only provide its numerical 
values at the two scales used in our illustrations below,
\beq
\label{c0}
  c^{}_0(1.956) \:\: = \:\: 88.28   \;\; , \quad
  c^{}_0(19.56) \:\: = \:\: 70.23   \;\; .
\eeq
For $F_2^{\,c}$, hence $\nf=3$ light flavours, our corresponding new NNLO 
results are numerically given~by

\pagebreak

\vspace*{-10mm}
\bea
\label{c2thresh}
  c_{2,g}^{\,(2)}(\xi, \beta) &\:\: \simeq \:\:&
  \frac{c_{2,g}^{(0)}}{(4\pi)^4} \:
  \Big\{
    \ln^4\beta \: 1152
    \: - \:
    \ln^3\beta \, \big( 1545. +1152\, L \big)
 \nn \\[-1.3mm] & & \mbox{\hspp} 
    \: + \:
    \ln^2\beta \, \Big( 
      - 3570. + 48\,c^{}_0(\xi)
      + ( 118.0 + 48\,\bar{c}^{}_0(\xi) )\, L + 288 L^2 
      - 16\, \pi^2 \beta^{-1} \Big)
\nn \\[-0.3mm] & & \mbox{\hspp}
    \: + \: 
    \ln \beta \, \Big( 2403.  - 20.19\,c^{}_0(\xi) 
      + \big( 2223. - 20.19\, \bar{c}^{}_0(\xi) - 24\,c^{}_0(\xi) \big)\, L 
\nn \\[-0.3mm] & & \mbox{\hspp\hspp}
    + \big( 291.3\, - 24\,\bar{c}^{}_0(\xi) \big) L^2 
    + \pi^2 \beta^{-1} [ 2.910 + 8\,L ] \Big)
    \;\; + \;\; O(\beta^{-2}) 
 \Big\} \:\: \quad
\eea
with $L \equiv \ln (\mu^2/m^2)$,
where the coefficients with a decimal point are approximate. In addition to the
terms given here, also the non-logarithmic $1/\beta$ Coulomb contributions are 
now known.

\begin{figure}[thb]
\centerline{\epsfig{file=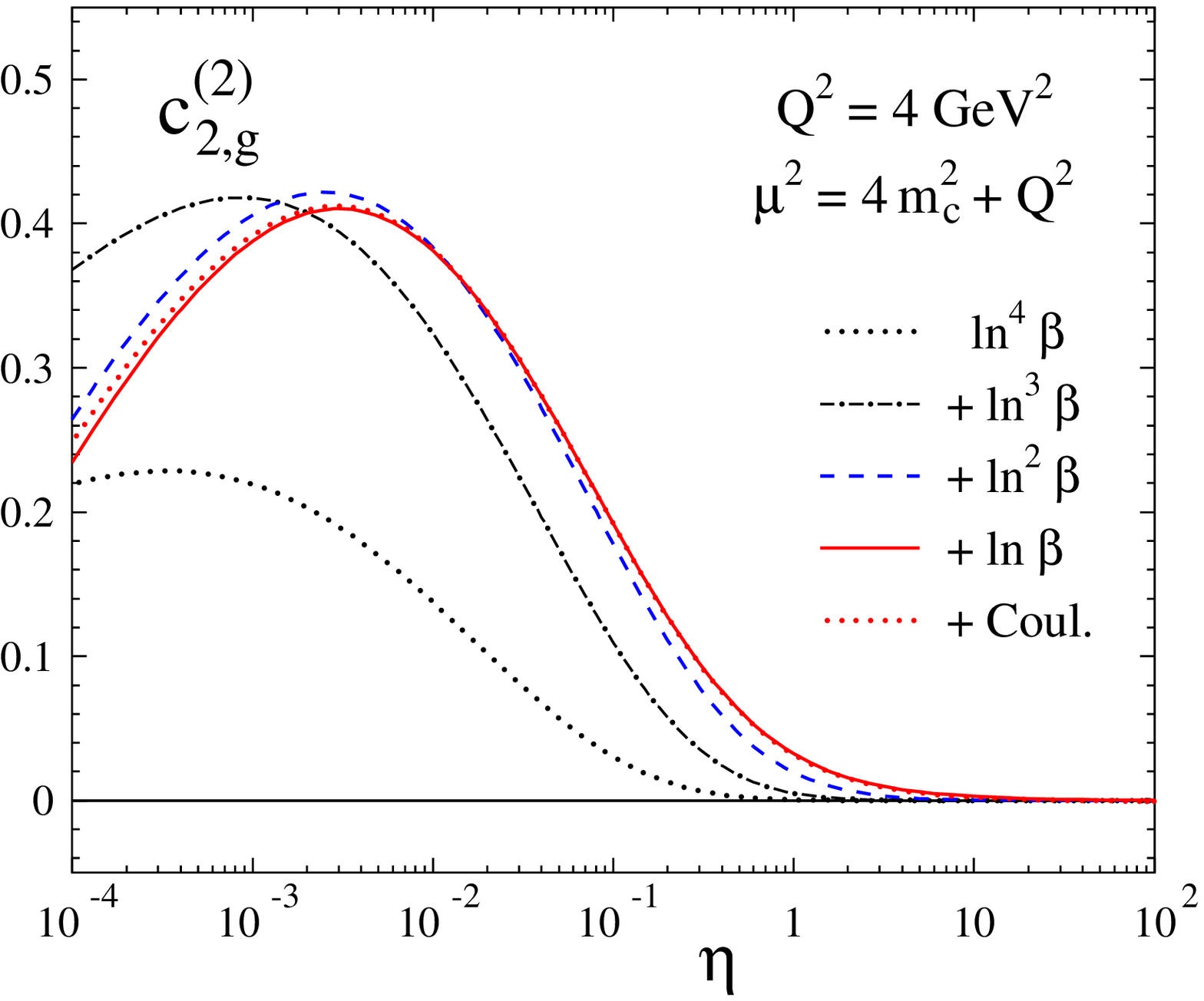,width=6.5cm,angle=0}
\epsfig{file=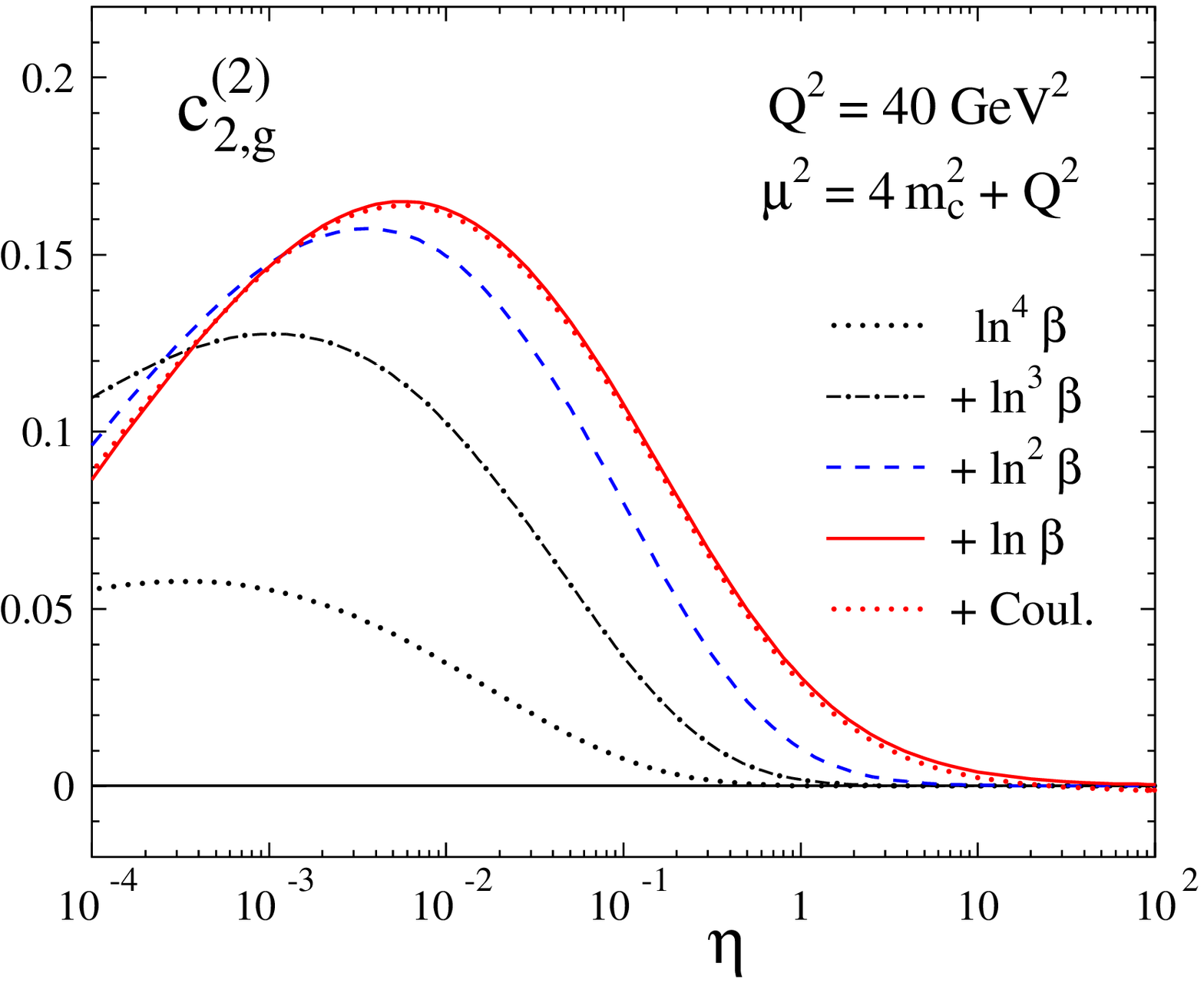,width=6.5cm,angle=0}}
\vspace{-2mm}
\caption{
 Successive approximations to the NNLO gluon coefficient function for
 $F_2^{\:\!c}$ in terms of threshold logarithms and $1/\beta$ Coulomb
 contributions at two typical scales $\Qs$ for a charm pole-mass $m = 1.43$ GeV.
 }
\vspace{-1mm}
\end{figure}

The threshold expansion (\ref{c2thresh}) of the NNLO coefficient function is 
shown in Fig.~1 for a standard choice of the renormalization$/$factorization 
scale $\mu$. Keeping only the highest two logarithms is obviously insufficient. 
The new $\ln \beta$ contribution is rather small at the lower, but definitely 
relevant at the higher scale, while the non-logarithmic NNLO Coulomb terms are 
small in both cases.
The resulting estimates for the NNLO corrections to $F_2^{\,c}$ are illustrated
in Fig.~2. In the region $10^{-4} \lsim x \lsim 10^{-2}$ these amount to no 
more than about $5 - 10\%$ at $Q^2\,=\,40\:\GeVs$, but reach $15 - 30\%$ at 
$Q^2\,=\,4\;\GeVs$, which the largest effects occurring at the upper end of the 
above $x$-range.

\begin{figure}[bth]
\centerline{\epsfig{file=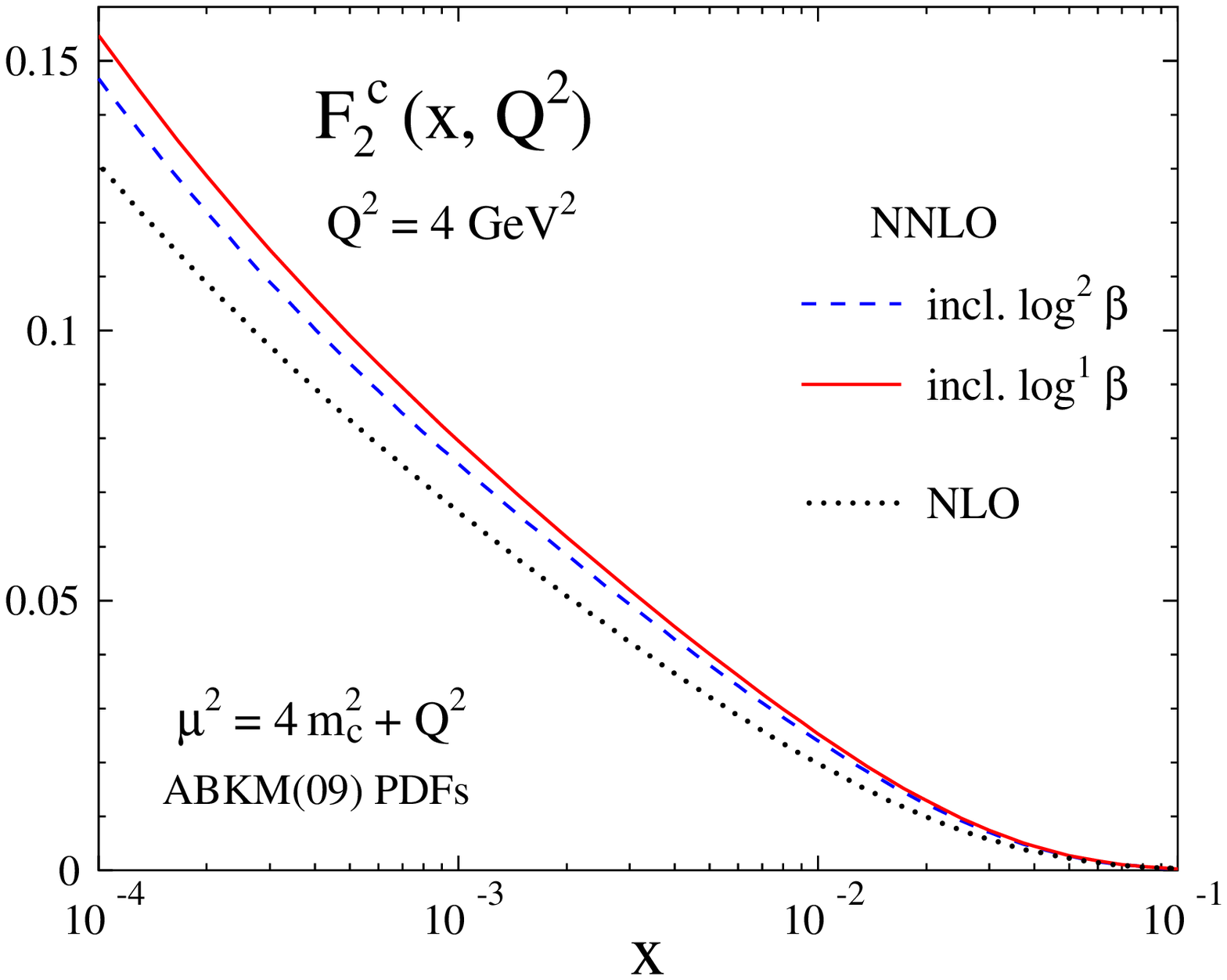,width=6.5cm,angle=0}
\epsfig{file=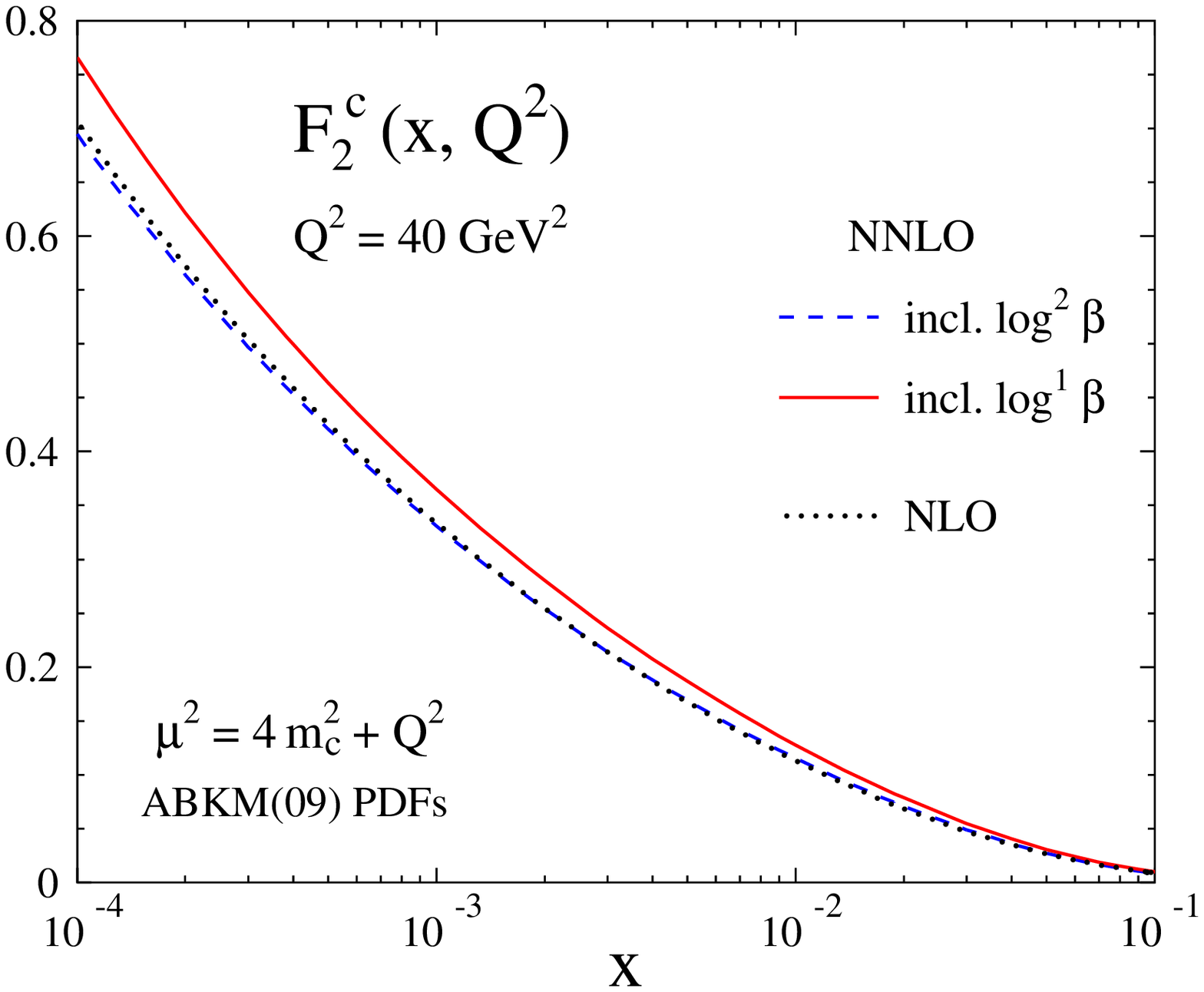,width=6.5cm,angle=0}}
\vspace{-3mm}
\caption{
 NLO and threshold-estimated NNLO results for the charm contribution to the
 structure function $F_2$ using the respective parton distributions and strong 
 coupling constants of Ref.~\cite{ABKM09} with $m_c = 1.43$ GeV.
 }
\vspace{-5mm}
\end{figure}

\section{The \boldmath $p^{}_T$-differential charm structure function}

\noindent
Experimentally the inclusive structure function $F_2^{\,c}$ is determined via
(theory-dependent) extrapolations of more differential cross sections.
As an example we consider the $p^{}_T$-unintegrated structure function 
$dF_2/dp^{}_T$, calculated at NLO in Refs.~\cite{LRSvN93a}. 
First NNLO estimates based on the next-to-leading logarithmic (NLL) threshold
resummation were derived in Ref.~\cite{LM98}. In Fig.~3 we present an update
of these predictions, using an independent code and up-to-date parton
densities \cite{ABKM09}.  
	
\begin{figure}[thb]
\centerline{\epsfig{file=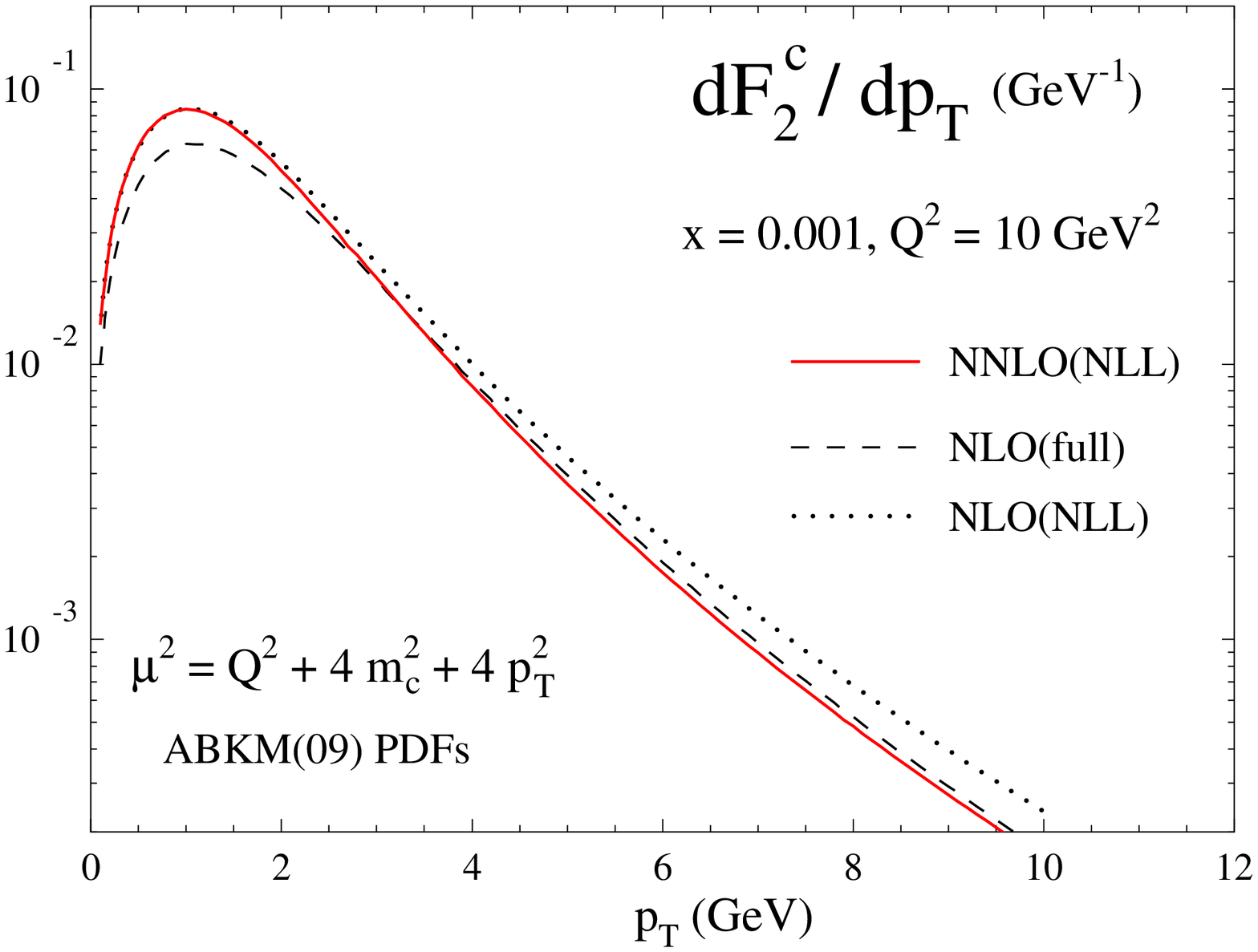,width=7.0cm,angle=0}
\epsfig{file=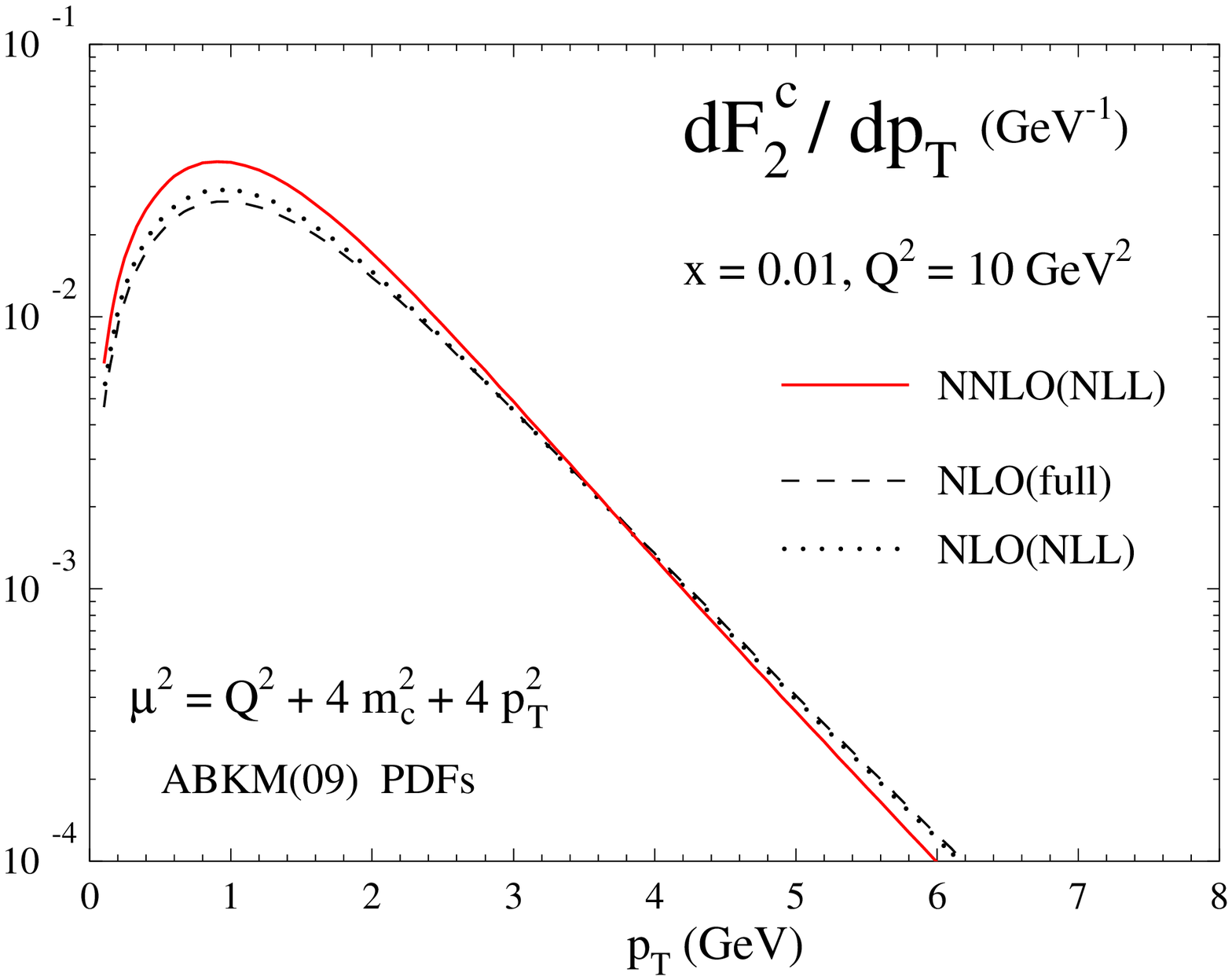,width=7.0cm,angle=0}}
\vspace{-2mm}
\caption{NNLO estimates for the $p^{}_T$-unintegrated charm structure function
$F_2$ for two typical values of $x$. At NLO the results for the NLL 
expanded coefficient function are compared compared to the exact values.
 }
\end{figure}

The NLO comparison of the complete and NLL expanded results indicates that the
latter are reliable at $x \simeq 0.01$, but not at $x \simeq 0.001$. 
The estimated NNLO corrections are large and positive around the peak of the
distribution, where they amount to as much as 40\%. 
More work is needed to arrive at quantitatively reliable NNLO predictions for
this and other differential cross sections. It is interesting to note, however,
that a considerable excess over the NLO results has been observed in HERA 
measurements of charm production including very low values of $p_T^{}$ 
\cite{ZEUS10ch}.

\section{Summary and Outlook}

\noindent
We have determined the next-to-next-to-leading logarithmic (NNLL) resummation 
exponent and the one-loop matching function for the dominant gluon coefficient 
function for the heavy-quark structure functions $F_2^{\,h}$ in deep-inelastic
lepton-hadron scattering. The results have been used to obtain all 
threshold-enhanced NNLO contributions to this coefficient function, which we
have illustrated for the especially important case of charm production.

\vspace{1mm}
At present, these results provide the only reliable estimate of the NNLO 
effects at small scales, $Q^2 \gsim \hspace*{-3mm} / \; 10\: m_c^{\:\!2}$. 
At larger scales, it may be useful to combine these threshold contributions, 
the Mellin moments (with respect to $x$) of the large-$\xi$ limits~\cite{BBK09}
and the leading large-$\eta$ (small-$x$) logarithms \cite{CCH91}, in order to 
obtain an all-$\eta$ approximate NNLO coefficient function. 

\vspace{1mm}
As an example for less inclusive quantities, we have also presented NNLO
threshold estimates for the $p_T^{}$-differential structure function. Also 
here the accuracy reached in Ref.~\cite{LM98} needs to be improved for quantitatively reliable predictions. Present results indicate considerably
larger NNLO corrections than for $F_2^{\,c}$ close to the peak of the
distribution at rather low values of~$p_T^{}$. 
	

\end{document}